\documentstyle[PASJadd,psfig]{PASJ95}
\newcommand{\vol}{}
\newcommand{\no}{}
\newcommand{\lett}{}
\newcommand{\spage}{}
\newcommand{\rdate}{}
\newcommand{\adate}{}

\begin{document}
\setcounter{page}{100}

\title{The compact core-jet region of the superluminal quasar 3C\,216}

\author{
Zsolt {\sc Paragi}, S\'andor {\sc Frey} and Istv\'an {\sc Fejes} \\
{\it F\"OMI Satellite Geodetic Observatory, Penc, Hungary} \\
{\it E-mail(ZP): paragi@sgo.fomi.hu}  \\[6pt]
Tiziana {\sc Venturi} \\
{\it Istituto di Radioastronomia, Bologna, Italy} \\[6pt]
Richard W.\ {\sc Porcas} \\
{\it Max-Planck-Institut f\"{u}r Radioastronomie, Bonn, Germany} \\
and \\
Richard T.\ {\sc Schilizzi} \\
{\it Joint Institute for VLBI in Europe, Dwingeloo, The Netherlands} \\
{\it Leiden Observatory, Leiden, The Netherlands} \\
}

\abst{Space very long baseline interferometry (VLBI) observations of 
the quasar 3C\,216 with the VLBI Space Observatory Programme
(VSOP) reveal that the parsec scale structure of the source can be well
described by compact jet models developed for interpreting the core region 
of radio-loud active galactic nuclei (AGN). The measured brightness 
temperature of $T_{\mathrm B}=7.9 \times 10^{11}$\,K is comparable to the
inverse Compton limit, from which we determine a lower limit of 
$\delta \sim 3.17$ for the Doppler-boosting factor. The apparent transverse
velocity of the superluminal component is 
$\beta_{\mathrm{app}}=(3.0\pm 0.2)\, h^{-1}$ assuming a constant velocity,
but deceleration of the jet material cannot be excluded from our data.
A combination of the above values indicates that the viewing angle of the 
core-jet to the line-of-sight is less than 
$18^{\circ}\hspace{-4.5pt}.\hspace{.5pt}4$, and the jet 
Lorentz-factor exceeds 3.16\,. The observed small size of the source is
probably caused by both interaction with the interstellar medium, and a
projection effect.}

\kword{Astrophysical jets --- Galaxies: active --- Quasars: individual (3C\,216)}

\maketitle
\thispagestyle{headings}

\section
{Introduction}

The central region of radio-loud active galactic nuclei (AGN) can
be studied with sub-milliarcsecond (mas) resolution using the space 
very long baseline interferometry (space VLBI) technique; 
this corresponds to only a few parsecs projected linear 
size even for the highest redshift sources. The inclusion of an orbiting
element into the VLBI array is important because the achievable resolution
increases without using higher observing frequencies. As theory predicts
that the size of the optically thick region of compact jets is inversely
proportional to the frequency (Blandford \& K\"onigl 1979), sources that are 
too compact for ground VLBI arrays can only be resolved by space VLBI. 
The other advantage of this technique over ground-based VLBI is that higher 
brightness temperatures can be measured. The maximum achievable brightness
temperature (for an isotropic source) is limited by inverse Compton scattering
to $T_{\rm B}=10^{12}$~K (Kellermann \& Pauliny-Toth 1969); 
$T_{\rm B}$ values exceeding $\sim 10^{12}$~K indicate Doppler-boosting of 
the emission. Observational evidence for Doppler-boosting and investigation of 
parsec-scale jet misalignments at sub-mas resolution (e.g. Tingay et al.\ 1998)
help us to refine orientation-based unification models.

Below we present results of our VLBI Space Observatory Programe (VSOP)
observations of 3C\,216 (0906+430). 
The source ($z=0.67$, Spinrad et al.\ 1985) shows high optical polarization
and variability, typical of the class of blazars (Angel \& Stockman 1980).
It was classified as a compact steep spectrum (CSS) quasar as well 
(e.g. Fanti et al.\ 1985; Pearson \& Readhead 1988), though this classification 
may not be entirely correct as the observed small source size ($\sim$15~kpc) 
is probably not intrinsic, and within $\sim 100$ mas of the centre the radio 
structure is reminiscent of core dominated sources (Fejes et al.\ 1992). 
There are slightly misaligned radio lobes on arcsecond scales oriented 
to South-West and North-East directions from the centre 
(Pearson et al.\ 1985; Fejes et al.\  1992); the latter seems to align with 
the barely-resolved optical structure observed by the Hubble Space Telescope 
(de Vries at al.\ 1997). Images made by the VLA at 1.4~GHz (Barthel et al. 1988)
and by the MERLIN array at 408~MHz (Fejes et al.\ 1992) reveal the presence of 
a low surface brightness halo about 8" in size -- there is no evidence for
larger scale emission at lower frequencies, according to 81.5~MHz long 
baseline interferometric observations (Hartas et al. 1983).
Regardless of the strong lobes that dominate 
the spectrum at low frequencies, 3C\,216 shows the structural properties 
of a core dominated source at centimeter wavelengths on VLBI scales.
It has a bright core component and a jet that is perpendicular to
to the arcsecond-scale radio lobes (Fejes et al.\  1992). The jet follows a gently 
curved path, with underlying wiggles, and ends in a sharp bend at 140~mas from
the centre (Fejes et al.\  1992; Akujor et al.\  1996).
In the innermost part of the source, components emerge at superluminal 
speeds (Barthel et al.\  1988). Venturi et al.\  (1993) showed
that deceleration takes place in the inner few mas (10-20~parsecs). 
There is also a quasi-stationary component at $\sim 1.5$ mas from the 
radio core at 5~GHz. Our main goal was to investigate this region in
more detail.

\begin{table}[ht]
\small
\begin{center}
Table~1.\hspace{4pt} Characteristics of radio antennas used in our VSOP 
experiment. \\
\end{center}
\vspace{6pt}
\begin{tabular*}{\columnwidth}{@{\hspace{\tabcolsep}
\extracolsep{\fill}}p{7pc}rr} 
\hline\hline\\[-6pt]
Name                        & Diameter (m)      & SEFD$^{\rm a}$ (Jy) \\   
[4pt]\hline \\[-6pt]


Effelsberg\dotfill          & 100               &     20 \\  
Westerbork\dotfill          & 93$^{\rm b}$      &     70 \\  
Jodrell Bank\dotfill        & 25                &    320 \\  
Noto\dotfill                & 32                &    260 \\  
Onsala$^{\rm c}$\dotfill    & 25                &    600 \\  
Torun\dotfill               & 32                &    250 \\  
Medicina\dotfill            & 32                &    296 \\  
Green Bank\dotfill          & 43                &    120 \\  
HALCA\dotfill               & 8                 &  15300 \\  

\hline
\end{tabular*}
\begin{list}{}{}
\item[$^{\rm a}$] System Equivalent Flux Density. \
\item[$^{\rm b}$] Used in phased array mode, an equivalent diameter is given. \
\item[$^{\rm c}$] Did not observe due to high winds. \
\end{list}
\end{table}

\section
{Observations, data reduction and imaging}

The source was observed at 5~GHz with the array of the VSOP satellite HALCA, 
the 43m Green Bank telescope and the western part of the European VLBI 
Network (EVN) on 14/15 February 1999. Characteristics of the radio antennas 
are listed in Table~1. There were three tracking passes during
the 8.5 hour experiment using the Green Bank, Tidbinbilla and Goldstone 
tracking stations. The data were recorded in left circular polarization
in VLBA mode, and correlated at the NRAO VLBA correlator in
Socorro (USA). We used AIPS (Cotton 1995; Diamond 1995) version 15OCT98 for
data processing at the F\"OMI Satellite Geodetic Observatory, Hungary. 
The measured antenna system temperatures ($T_{\mathrm{sys}}$) were used for 
a-priori amplitude calibration where available. 
We used nominal $T_{\mathrm{sys}}$ values for Green Bank and HALCA.
Fringes were found with high signal-to-noise ratio for space baselines 
during the experiment, except in the short third tracking pass. We averaged
data in frequency and time (one minute integration time) and performed 
imaging in DIFMAP (Shepherd et al. 1994).


\begin{center}
\begin{figure}
\vspace{90mm}
\includegraphics{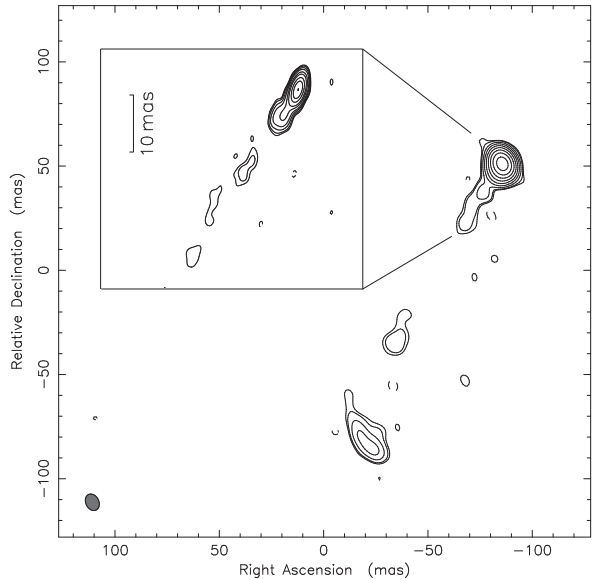}
\caption{Contour representation of the bending jet of 3C\,216, imaged using
natural
weighting of the data taken at 5~GHz. Due to the large image pixel size of one
mas, only data from the shorter EVN baselines were included. Contour levels are
$-$0.35, 0.35, 0.5, 1, 2, 4, 8, 16, 32, 64\% of the peak brightness of
507~mJy/beam, the restoring beam is $8.3 \times 6.3$~mas at $PA=27^{\circ}$. 
The inset shows the wiggling jet reproduced with 0.4 mas image cellsize and some
$uv$-tapering, which resulted in a restoring beam of $3.0 \times 1.4$~mas at 
$PA=-15^{\circ}$ (about the resolution of the ground-only array of this
experiment). Contour levels are $-$0.3, 0.3, 0.5, 1, 2, 5, 10, 25, 50, 99\% of 
the peak brightness of 461~mJy/beam.}   
\label{fig1}
\end{figure}
\end{center}


\begin{center}
\begin{figure*}
\vspace{80mm}
\includegraphics{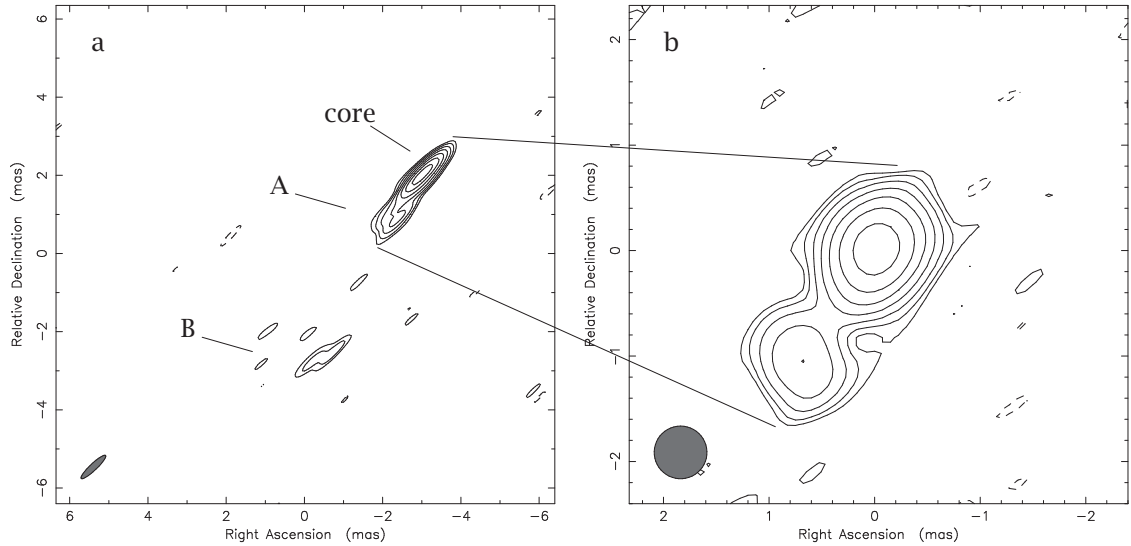}
\caption{Full resolution space VLBI image. Contour levels are $-$1, 1, 
2, 4, 8, 16, 32, 64\% of the peak brightness of 302~mJy/beam, the parameters
of the restoring beam are $0.86 \times 0.17$~mas, $PA=-47^{\circ}$. 
{\bf b} The core region with a half milliarcsecond restoring beam.
The contour levels are the same as in {\bf a}, the peak brightness
is 319~mJy/beam.}   
\label{fig2}
\end{figure*}
\end{center}

We experimented with a variety of $uv$ cut-offs and data weighting in imaging,
in order to image the source structure on different spatial scales. We were 
able to reconstruct the jet bend at $\sim 140$~mas from the core using natural 
weighting 
-- this feature was resolved out using uniform weighting,
when ground-space baselines dominate in imaging. We found that the gain of
the EVN telescopes -- which have baselines sensitive to the 100 mas scale 
structure -- could be better self-calibrated if we added clean components from 
the large scale emission "manually", when performing imaging with uniform 
weighting. 

The process is described below. 1) A self-calibration and imaging cycle was
carried out using natural weighting. 2) The source model file obtained was
edited so that clean components within 5~mas from the phase centre
were deleted (emission from larger scales do not have a significant contribution
to the correlated flux density on the Earth-space
baselines). 3) A new self-calibration and imaging cycle was started with uniform
weighting of the data. After the starting phase self-calibration to a 1~Jy point
source model, the edited model file from 1) was appended to the clean component model.
Imaging of the inner few mas region of the source was then 
continued in the usual way. This method should be applied with caution but proved 
to be useful in our case. While the basic structure of the source 
obtained as described above was not different from what can be achieved
by the standard way, we were able to improve the dynamic range of the final 
space VLBI map. 

\section
{Results}

Contour representations of the resulting images -- according to the various 
imaging strategies -- are shown in Fig.~1. and Fig.~2. 
The bright knot at the jet bend can be identified at 140~mas from the core
in Fig.~1., where the inset shows the wiggling jet in more detail.
Components in the full resolution space VLBI image (Fig.~2a) are 
labelled in the same way as in Venturi et al.\  (1993). The core region is
resolved; however, the clean beam is highly elongated roughly along the jet
direction. The results of model fitting of the self-calibrated data are listed 
in Table~2. Using a 0.5~mas circular restoring beam, Component A is apparently 
well separated from the core (Fig.~2b). The separation is estimated as
1.25$\pm$0.05~mas. This is not in agreement with the results of model 
fitting; an elongated component closer to the core (0.87~mas) gives 
the best fit to the data (see Table~2.). Component B is clearly 
resolved, and seems to be edge brightened on the western side. 
It is difficult to estimate the positional uncertainty of component B; 
we will use 0.2~mas in further analysis (as well as for other epochs). 
The position angle of the inner jet components ($PA=152^{\circ}$) is 
slightly misaligned with respect to the 100~mas scale jet
($PA=146-149^{\circ}$). 

In order to scale angular sizes to projected linear sizes, we used
$H_{0}=100\,h\,$km\,s$^{-1}$\,Mpc$^{-1}$ and $q_{0}=0.5$;  
1~mas corresponds to $3.9\,h^{-1}$pc at the distance of 3C\,216. 
We also calculated the observed brightness temperatures of the source 
components in the core region, after correction for the $z$ cosmological
redshift, using the formula given by Kellermann and Owen (1988): 

\begin{equation}
T_{\mathrm{B}}[K] = 1.22 \times 10^{12} \,\, (1+z) \, \frac{S}{\theta_{1} \theta_{2} \nu^{2}} \; , 
\end{equation}
where $S$ is the flux density in Jy, $\theta_{1}$ and $\theta_{2}$ are the 
sizes of the major and minor axes of the fitted Gaussian component in mas, 
and $\nu$ is the observing frequency in GHz. These values are listed in 
Table~2. For the radio core, $T_{\mathrm B}=7.9 \times 10^{11}$\,K, 
which is comparable to the inverse Compton limit 
(Kellermann \& Pauliny-Toth 1969).


\section{Discussion}

The overall radio structure of 3C\,216 can be modelled with either helical 
flows or precessing beams in which jet components move ballistically, but a
unique solution cannot be found (Fejes et al.\ 1992). Helical jets are proposed
to exist in some radio-loud AGN (e.g. Conway \& Murphy 1993, and references
therein) e.g. due to Kelvin-Helmholtz instabilities introduced by a driving
mechanism such as precession of the central engine. Note that precessing jet
models invoking ballistic ejection of plasmons are less favoured at present as
it would require two orders of magnitude higher density of the jet with respect
to the ambient medium, which is not observed in AGN (Hardee et al.\ 1994). 

In the case of 3C\,216, Akujor et al. (1996) noted that the edge brightened
structure and the flattened spectrum of the component near the jet bend may
indicate interaction with the interstellar medium. Such jet-ISM interaction 
is proposed to be a general phenomenon in the sample of CSS sources (including
3C\,216) investigated by de Vries et al. (1999) with the Hubble Space 
Telescope, which gives evidence for the presence of emission-line gas aligned
with the radio structures. The very high rotation measure observed in the 
jet bend (Venturi \& Taylor 1999) and the recently discovered H\,{\sc i} 
absorption in the source  (Pihlstr\"om et al.\  1999) indicate that the jet is 
indeed deflected by an interstellar cloud in the host galaxy. High resolution 
H\,{\sc i} observations have been carried out in the UHF band with the EVN to
identify the location of the absorbing gas (Y. Pihlstr\"om, priv. comm.).

\begin{table*}
\small
\begin{center}
Table~2.\hspace{4pt}Fitted elliptical Gaussian model parameters of the source 
components.
$S$: flux density, $r$: separation, $\Theta$: position angle, $a$: major axis,
$b/a$: ratio of minor and major axes, $\Phi$: position angle of the major axis,
$T_{\mathrm B}$: brightness temperature. \\
\end{center}
\vspace{6pt}
\begin{tabular*}{\textwidth}{@{\hspace{\tabcolsep}
\extracolsep{\fill}}p{7pc}rrrrrrr} 
\hline\hline\\[-6pt]
Component 	& $S$    & $r$    & $\Theta$  & $a$\,\,\,\,  & $b/a$ & $\Phi$      & $T_{\rm B}$ \\
          	& (mJy)  & (mas)  &($^{\circ}$)& (mas)       &       &($^{\circ}$) & ($10^{12}$ K) \\
[4pt]\hline \\[-6pt]

Core\dotfill	& 381.6  & 0.00   & --        & 0.36         & 0.31  & $-$32       & 0.79  \\ 
A\dotfill	& 122.8  & 0.87   & 151.5     & 1.37         & 0.21  & $-$62       & 0.03  \\ 
B\dotfill	& 45.1   & 5.16   & 152.3     & 1.93         & 0.14  & $-$31       & $<$0.01  \\

\hline
\end{tabular*}
\end{table*}


\begin{center}
\begin{figure}
\vspace{70mm}
\includegraphics{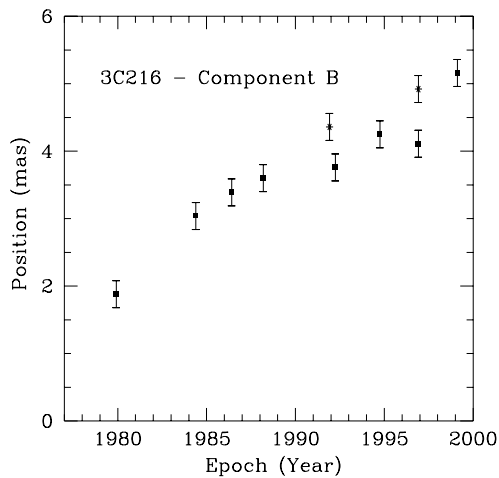}
\caption{Motion of the superluminal component B from the earliest VLBI
observations. The 1979--89 data are taken from Barthel et al.\  (1988) and
 Venturi et al.\  (1993), the 1991--98 data are taken from Venturi et al.\  
(in prep.). Filled rectangles indicate observatons at 5~GHz, while asterisks 
at 8.4~GHz.}   
\label{fig3}
\end{figure}
\end{center}

The position of component B from earlier VLBI measurements is shown 
in Fig.~3. These data are consistent with a constant proper 
motion of $\mu=0.14\pm 0.01$~mas/yr between 1979 and 1999.
But note that the higher resolution observations (the two 8~GHz
data points and our 5~GHz space VLBI measurement) show significantly
larger separations of the component from the centre. 
The ground only data points obtained at 5~GHz are still indicative of
jet deceleration or orientation changes, if we compare the trend in the 
component separation evolution before and after 1988. It may be misleading 
to compare component separations measured with instruments of different
resolution or measured at different frequencies, which may result in 
a shift of the reference point (the location of the observed brightness
peak). This can happen in a case e.g. when a jet component is blended
with the core in a lower resolution observation. But the reference
point is also expected to shift with the observing frequency due to a change in
synchrotron self-absorption opacity, if the core region can be described with 
compact jet models 
(Lobanov 1988 and referencies therein, see below). Such a shift was recently 
detected e.g. in the BL~Lac objects 1803+784 (P\'erez-Torres et al. 2000) and 
1823+568 (Paragi et al. 2000). The latter was observed at four-frequencies 
(5--22 GHz) with the VLBA; these observations show a systematic increase of 
core--jet component separation with increasing frequency.

The topic of constant component speed vs. jet deceleration scenario for 3C\,216 will 
be investigated further by T. Venturi et al. (in preparation). For the present discussion we 
assumed the most simple case of constant component speed, and calculated the apparent 
transverse velocity with the following formula (e.g. Pearson \& Zensus (1987)):

\begin{equation}
\beta_{\mathrm{app}} = \mu \, \frac{z}{H_{0}(1+z)} \, 
\left[ \frac{1+(1+2q_{0}z)^{1/2}+z}{1+(1+2q_{0}z)^{1/2}+q_{0}z} \right].
\end{equation}
The observed proper motion corresponds to $\beta_{\mathrm{app}}=(3.0\pm 0.2)\, h^{-1}$ 
apparent transverse velocity.

The nature of component A is rather unclear. Its separation from the core 
is 1.25~mas (see the super-resolved image in Fig.~2b), compared to 
the earlier reported quasi-stationary position of 1.4--1.6~mas 
(Venturi et al.\  1993). The beam orientation is not suitable 
to decide whether this feature is a separate component, or just
an extension to the core emission. As we could not model the source 
structure with a stable component at a distance of 1.25~mas 
in the model-fitting process, the latter seems more favourable. It is 
very probable that the "core" and "component A" together form the 
so-called ultracompact jet region of 3C\,216 (see Lobanov 1998). 
In this case we can apply the equipartition jet model of 
Blandford and K\"onigl (1979) in which the limiting brightness 
temperature is about $3 \times 10^{11}\, \delta^{5/6}\,$K.  
A comparison with the observed value results in a lower limit to 
the Doppler factor: $\delta \sim 3.17$. The Lorentz factor of the 
jet ($\gamma$) and the viewing angle to the line of sight 
($\theta$) can be given by (e.g. Daly et al.\ 1996):

\begin{equation}
\gamma = (2 \delta)^{-1}\,(\beta_{\mathrm{app}}^{2}+\delta^{2}+1),
\end{equation}
and
\begin{equation}
\cos \theta = \left( \frac{1}{\sqrt{\gamma^{2}-1}} \right) \, (\gamma - \delta^{-1}).
\end{equation}
Using the calculated values of $\beta_{\mathrm{app}}$ and the lower 
limit of $\delta$, and assuming $h = 0.7$ our results are
$\gamma > 3.16$ and $\theta < 18^{\circ}\hspace{-4.5pt}.\hspace{0.5pt}4$. 
Much higher Doppler factors for 3C\,216 have been determined by other methods
($\delta=33$ in Ghisellini et al.\  1992 and $\delta=65$ in G\"{u}ijosa \& Daly 1996).
These Doppler factors would result in very large Lorentz factors 
of $\gamma > 16$ and much smaller angles to the line of sight ($\sim 1^{\circ}$),
unless $\beta_{\mathrm{app}}$ represents a jet pattern speed instead of
bulk motion. Note that the $\delta$ estimations from the literature mentioned above
made use of VLBI data taken during an outburst, when the flux density of the source 
core was much higher (cf. Barthel et al. 1988).  

\section{Conclusions}

The blazar type core of the source can be interpreted in terms of compact jets. 
We estimate that the inner jet is oriented close to the line of sight with
$\theta < 18^{\circ}\hspace{-4.5pt}.\hspace{0.5pt}4$, and the radio emission 
is Doppler-boosted by a factor of $\sim 3$ at the epoch of observations. We
conclude that the observed small projected size of 3C\,216 is probably caused 
by both interaction of the radio beams with the interstellar medium and a
projection effect. The nature of the core region should be investigated in more
detail in future high resolution multi-frequency VLBI experiments, which would
allow us to compare the observed spectral distribution with physical models.

\vspace{1pc}\par

We gratefully acknowledge the VSOP Project, which is led by the Japanese 
Institute of Space and Astronautical Science in cooperation with many 
organizations and radio telescopes around the world. This research was 
supported in part by the Netherlands Organization for Scientific Research 
and the Hungarian Scientific Research Fund (grant no. N31721 \& T031723),
as well as the Hungarian Space Office. The National Radio Astronomy 
Observatory is operated by Associated Universities, Inc. under a Cooperative
Agreement with the National Science Foundation. We thank the anonymous referee
for useful comments.

\section*{References}
\small


\re 
Angel J.R.P., Stockman H.S.\ 1980, ARA\&A 18, 321

\re
Akujor C.E., Porcas R.W., L\"udke E., Shone D.L.\ 1993, in
Davies E., Booth R.J. (eds.) Sub-arcsecond Radio Astronomy,
Cambridge University Press, p. 265

\re 
Akujor C.E., Porcas R.W., Fejes I.\ 1996, in  Ekers R., Fanti C., Padrielli L. 
    (eds.) Extragalactic Radio Sources, Dordrecht: Kluwer, p. 83

\re 
Blandford R.D., K\"onigl A.\ 1979, ApJ 232, 34

\re 
Barthel P.D., Pearson T.J., Readhead A.C.S.\ 1988, ApJ 329, L51 

\re
Conway J.E., Murphy D.W.\ 1993, ApJ 411, 89

\re 
Cotton W.D.\ 1995, in Zensus J.A., Diamond P.J., Napier P.J. (eds.)
Very Long Baseline Interferometry and the VLBA,
ASP Conference Series Vol. 82, p. 189

\re 
Daly R.A., Guerra E.J., G\"uijosa A.\ 1996, in 
Hardee P.E., Bridle A.H., Zensus J.A. (eds.) 
Energy Transport in Radio Galaxies and Quasars, 
ASP Conference Series Vol. 100, p. 73

\re 
Diamond P.J.\ 1995, in Zensus J.A., Diamond P.J., Napier P.J. (eds.)
Very Long Baseline Interferometry and the VLBA,
ASP Conference Series Vol. 82, p. 227

\re Fanti C., Fanti R., Schilizzi R.T., Spencer R.E., van Breugel W.J.M.\ 1985,
A\&A 143, 292

\re 
Fejes I., Porcas R.W., Akujor C.E.\ 1992, A\&A 257, 459

\re 
Ghisellini G., Celotti A., George I.M., Fabian A.C.\ 1992, MNRAS 258, 776


\re 
G\"{u}ijosa A., Daly R.A.\ 1996, ApJ 461, 600

\re
Hardee P.E., Cooper M.A., Clark D.A.\ 1994, ApJ 424, 126

\re
Hartas J.S., Rees W.G., Scott P.F., Duffett-Smith P.J.\ 1983, MNRAS 205, 625

\re 
Kellermann K.I., Owen F.N.\ 1988, in Verschuur G.L., Kellermann K.I. (eds.) 
Galactic and Extragalactic Radio Astronomy, Springer, p. 577

\re 
Kellermann K.I., Pauliny-Toth I.I.K.\ 1969, ApJ 155, L71

\re 
Lobanov A.P.\ 1998, A\&A 330, 79

\re
Paragi Z., Fejes I., Frey S.\ 2000, in Vandenberg N.R., Baver K.D. (eds.)
International VLBI Service for Geodesy and Astrometry 2000 General Meeting Proceedings,
NASA/CP--2000--209893

\re
Pearson T.J., Readhead A.C.S.\ 1988, ApJ 328, 114


\re
Pearson T.J., Zensus J.A.\ 1987, in Zensus J.A., Pearson T.J. (eds.) 
Superluminal Radio Sources, p. 1 

\re
Pearson T.J., Perley R.A., Readhead A.C.S.\ 1985, AJ 90, 738

\re
P\'erez-Torres M.A., Marcaide J.M., Guirado J.C. et al.\ 2000, A\&A 360, 161

\re 
Pihlstr\"om Y.M., Vermeulen R.C., Taylor G.B., Conway J.E.\ 1999, 
ApJ 525, L13

\re 
Shepherd M.C., Pearson T.J., Taylor G.B.\ 1994, BAAS 26, 987

\re 
Spinrad H., Djorgovski S., Marr J., Aguilar L.\ 1985, PASP 97, 932


\re
Tingay S.J., Murphy D.W., Edwards P.G.\ 1998, ApJ 500, 673 

\re 
Venturi T., Pearson T.J., Barthel P.D., Herbig T.\ 1993, A\&A 271, 65

\re 
Venturi T., Taylor G.B. 1999, AJ 118, 1931

\re
de Vries W.H., O'Dea C.P., Baum S.A. et al.\ 1997, ApJS 110, 191

\re
de Vries W.H., O'Dea C.P., Baum S.A., Barthel P.D.\ 1999, ApJ 526, 27

\label{last}
\end{document}